\documentclass[sigconf]{acmart}
\usepackage{mathtools}
\usepackage{breakurl}
\usepackage{makecell}
\usepackage{algorithm}
\usepackage{algpseudocode}
\usepackage{booktabs} %For formal tables
\usepackage{enumitem}
\usepackage{letltxmacro}
\usepackage{multirow,array}
\usepackage{subfig}
\usepackage{scalerel}
\usepackage{listings}
\usepackage{multicol}
\usepackage{balance}

\definecolor{gre}{RGB}{15, 140, 0}
\definecolor{evalOrange}{RGB}{239, 155, 0}

\definecolor{page1color}{RGB}{34, 185, 4}
\definecolor{page2color}{RGB}{128, 255, 104}
\definecolor{page3color}{RGB}{230, 230, 0}
\definecolor{page4color}{RGB}{109, 109, 109}
\definecolor{page5color}{RGB}{251, 0, 6}

\newcolumntype{.}{!{\vrule width 2pt}}
\newcolumntype{'}{!{\vrule width 1.5pt}}
\AtBeginDocument{%
  \providecommand\BibTeX{{%
    \normalfont B\kern-0.5em{\scshape i\kern-0.25em b}\kern-0.8em\TeX}}}

\setcopyright{acmcopyright}
\copyrightyear{2021}
\acmYear{2021}
\acmDOI{x.x/x.x}

\acmConference[WWW '21]{30th The Web Conference}{April 19--23, 2021}{Ljubljana, Slovenia}
\acmBooktitle{30th The Web Conference,
  April 19--23, 2021, Ljubljana, Slovenia}
\acmPrice{15.00}
\acmISBN{x-x-x-x-x/xx/xx}

\graphicspath{{Images/}}

\begin{document}
%credit: https://tex.stackexchange.com/a/346309
\settopmatter{printacmref=false} % Removes citation information below abstract
\renewcommand\footnotetextcopyrightpermission[1]{} % removes footnote with conference information in first column
\pagestyle{plain} 

\title{Garbage, Glitter, or Gold: Assigning Multi-dimensional Quality Scores to Social Media Seeds for Web Archive Collections}

\author{Alexander C. Nwala}
\affiliation{%
  \institution{Observatory on Social Media\\Indiana University}
  \city{Bloomington} 
  \state{Indiana} 
  \postcode{47405}
  \country{USA}
}
\email{anwala@iu.edu}

\author{Michele C. Weigle}
\affiliation{%
  \institution{Department of Computer Science\\Old Dominion University}
  \city{Norfolk} 
  \state{Virginia} 
  \postcode{23529}
  \country{USA}
}
\email{mweigle@cs.odu.edu}

\author{Michael L. Nelson}
\affiliation{%
  \institution{Department of Computer Science\\Old Dominion University}
  \city{Norfolk} 
  \state{Virginia} 
  \postcode{23529}
  \country{USA}
}
\email{mln@cs.odu.edu}

\renewcommand{\shortauthors}{Nwala et al.}
\pagestyle{empty}
\begin{abstract}
 From popular uprisings to pandemics, the Web is an essential source consulted by scientists and historians for reconstructing and studying past events. Unfortunately, the Web is plagued by link rot and content drift (reference rot) which causes important Web resources to disappear. Web archive collections help reduce the costly effects of reference rot by saving Web resources that chronicle important stories and events before they disappear. These collections often begin with URLs called seeds, hand-selected by experts or scraped from social media posts. The quality of social media content varies widely, therefore, we propose\footnote{This is an extended version of the ACM/IEEE Joint Conference on Digital Libraries (JCDL2021) paper.} a framework for assigning multi-dimensional quality scores to social media seeds for Web archive collections about stories and events. We leveraged contributions from social media research for attributing quality to social media content and users based on credibility, reputation, and influence. We combined these with additional contributions from the Web archive research that emphasizes the importance of considering geographical and temporal constraints when selecting seeds. Next, we developed the Quality Proxies (QP) framework which assigns seeds extracted from social media a quality score across 10 major dimensions: $popularity$, $geographical$, $temporal$, \textit{subject expert}, $retrievability$, $relevance$, $reputation$, and $scarcity$. We instantiated the framework and showed that seeds can be scored across multiple QP classes that map to different policies for ranking seeds such as prioritizing seeds from local news, reputable and/or popular sources, etc. The QP framework is extensible and robust; seeds can be scored when a subset of the QP dimensions are absent. Most importantly, scores assigned by Quality Proxies are explainable, providing the opportunity to critique them. Our results showed that Quality Proxies resulted in the selection of quality seeds with increased precision (by $\sim$0.13) when novelty is and is not prioritized. These contributions provide an explainable score applicable to rank and select quality seeds for Web archive collections and other domains that select seeds from social media.
\end{abstract}


\maketitle
\sloppy
\section{Introduction and Background}
\label{intro}

\begin{figure*}
  \includegraphics[width=0.98\textwidth]{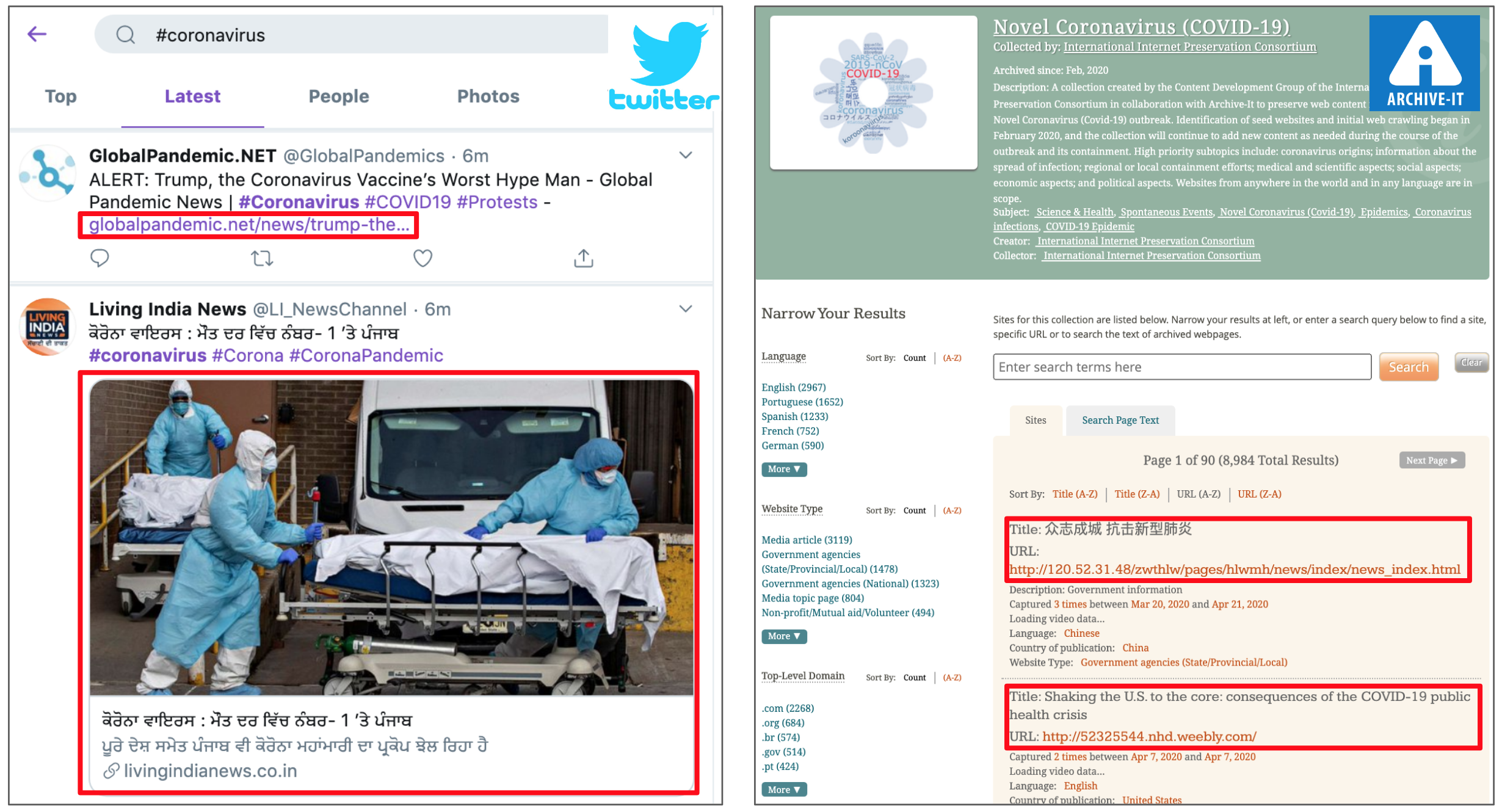} %
 
  \caption{Web archive collections such as the \textit{Coronavirus} collection (right image) begin with URLs called seeds or seed URLs (red annotations) often scraped from social media posts (left image)}
  \label{fig:twitterArchiveIt}
\end{figure*}

\begin{table}[]
\centering
\caption{Summary of the \textbf{Quality Proxies} (QPs) framework for assigning multi-dimensional QP scores to social media seeds for Web archive collections. A single QP belonging to either $popularity$, $proximity$, or $uncategorized$ class, assigns some quality trait to a seed, while different combinations of QPs combine to express different notions of quality (post popularity - $\{rp, lk, sh\}$, reputation - $\{re_b\}$, local authority - $\{re_n, ge_d\}$, etc.). For each QP we provided an instantiation (Section \ref{sec:qpFramework}) that approximates the QP class. The transparency and flexibility of the QP framework allows the user to extend and/or replace the instantiations of the QPs.}
\begin{tabular}{|c|c|c|c|}
\hline
\textbf{\#} & \textbf{Quality Proxy} & \textbf{Class} & \textbf{Acronym (s)} \\ \Xhline{4\arrayrulewidth}
1                                 & Posts                                        & Popularity                          & $rp$, $sh$, and $lk$ \\ \hline
2                                 & Author                                      & Popularity                          & $ap$                       \\ \hline
3                                 & Domain                                      & Popularity                          & $dp$                       \\ \Xhline{4\arrayrulewidth}
4                                 & Geographical-author                         & Proximity                           & $ge_a$                     \\ \hline
5                                 & Geographical-domain                         & Proximity                           & $ge_d$                     \\ \hline
6                                 & Temporal                                    & Proximity                           & $tp$                       \\ \hline
7                                 & Subject expert                              & Proximity                           & $su$                       \\ \hline
8                                 & Retrievability                              & Proximity                           & $rt$                       \\ \hline
9                                 & Relevance                                   & Proximity                           & $rl$                       \\ \Xhline{4\arrayrulewidth}
10                                & Reputation-broad                            & Uncategorized                       & $re_b$                     \\ \hline
11                                & Reputation-narrow                           & Uncategorized                       & $re_n$                     \\ \hline
12                                & Scarcity                                    & Uncategorized                       & $sc$                       \\ \hline
\end{tabular}
\label{tab:seedQPs}
\end{table}

On March 11, 2020, the World Health Organization declared the 2020 Coronavirus outbreak a pandemic. One week later, Archive-It (\textcolor{blue}{\url{https://archive-it.org/}}), an organization founded by the Internet Archive --- the largest public Web archive (\textcolor{blue}{\url{https://archive.org/}}) --- sent out a tweet\footnote{\textcolor{blue}{\url{https://twitter.com/archiveitorg/status/1240361850736381952}}} requesting social media users to contribute URLs about Coronavirus for preservation. It is important to preserve webpages chronicling important events such as the 2020 Coronavirus pandemic because according to SalahEldeen and Nelson, 11\% of Web resources shared on social media are lost after the first year of publication \cite{salaheldeen2012losing}, so we run the risk of losing a portion of our collective digital heritage if they are not preserved. The Internet Archive (IA) was founded in 1996, and since then, it has been archiving the Web by collecting and saving public webpages. This is based on a simple idea: an archived copy of a webpage may be viewed in place of a lost original copy, but this is only possible if the original webpage was saved. Archive-It is a service created by the Internet Archive where individuals and institutions create Web archive collections (e.g., Fig. \ref{fig:twitterArchiveIt}, right) that preserve webpages and their URLs, about a particular topic (e.g., \textit{2020 Coronavirus}). These collections begin with the selection of an initial list of URLs called seed URLs or seeds which are crawled as part of the preservation process called \textit{Web archiving}.

Following the occurrence of major news events, various organizations collect and save webpages about the events before they are lost due to reference rot \cite{KleinJCDL2018, zittrain2014perma, bar2004sic}. For example, archivists at the National Library of Medicine (NLM) saved seed URLs \cite{ChristieEbolaVirus} during the \textit{2014 Western African Ebola Virus Outbreak}. The NLM Ebola virus Web archive collection includes websites of organizations, journalists, healthcare workers, and scientists, related to the 2014 Ebola virus discourse. Similarly, archivists at Michigan State University saved webpages \cite{MSUFlint} chronicling the \textit{Flint Water Crisis} story. Unfortunately, we do not have enough curators to collect seeds amidst an abundance of local and global events, primarily because it is time-consuming to collect seeds manually, and collecting quality seeds requires domain expertise about the topic, which imposes an additional burden for curators. Consequently, organizations and researchers scrape social media posts for seeds (e.g., Fig. \ref{fig:twitterArchiveIt}, left) to cope with the shortage of curators with domain expertise \cite{yang2012study, priyatam2014seed}.
%\cite{twtReq0}
%\cite{twtReq0, twtReq1}
%\cite{KleinJCDL2018, zittrain2014perma, bar2004sic}
%\cite{yang2012study, priyatam2014seed}
While social media offers a cheap method of crowd-sourcing domain expertise, the quality of social media content varies widely. Selecting quality seed URLs from social media is challenging and has not been extensively studied in the Web archiving community, who acknowledges the importance of selecting good seeds but often pay more attention to the mechanisms of building collections. The challenge of selecting quality seeds is embodied in the idea that it is difficult to define ``quality,'' which could be subjective and is approximated with various metrics that are sensitive to relevance, popularity, or reputation. 

We developed the Quality Proxies (QPs) framework which generates a multi-dimensional quality score for seeds. A single QP assigns some quality trait to a seed, while different combinations of QPs combine to express different notions of quality (post popularity - $\{rp, lk, sh\}$, reputation - $\{re_b\}$, local authority - $\{re_n, ge_d\}$, etc.) that are used to score seeds and select those that exceed a user-defined threshold. The Quality Proxies framework was inspired by social media research for attributing quality to social media content and users based on credibility, reputation, and influence \cite{castillo2011information, pal2011identifying, canini2011finding}. It was additionally informed by Web archive research, which emphasizes the importance of geographical and temporal constraints when selecting seeds \cite{risse2014you, nwala2017local}, and consists of three main classes ($popularity$, $proximity$, and $uncategorized$) that sub-divide into additional classes enumerated in Table \ref{tab:seedQPs}.

Given that the Quality Proxies is a framework, we additionally instantiated the framework with metrics (Table \ref{tab:seedQPs}) that approximate each class. The transparency and flexibility of the framework means the user can extend it and/or replace a QP instantiation. The quality score of a seed can be assigned by extracting Quality Proxy metrics across all classes, or a subset of classes as input to a quality score function (Eqn. \ref{eqn:qpScore}), and selecting seeds that exceed a threshold.

Our contributions are as follows. First, Quality Proxies provide a flexible means of encoding multiple definitions of quality instead of a single definition of quality based on relevance or popularity. By providing a multi-dimensional framework, a curator can use the Quality Proxies to score and select not just popular (post popularity - $\{rp, lk, sh\}$) or relevant ($rl$) seeds but seeds from reputable sources ($re_b$), local news organizations ($ge_d$), popular residents of a local community ($\{ap, ge_a\}$), hard-to-find seeds ($\{rl, rt\}$), etc. This is because Quality Proxies independently behave like alphabets that can be combined in different ways to provide various policies for scoring seeds. Second, the QP framework is robust to enable the assignment of quality scores even when a subset of the metrics is absent. Third, we instantiated each QP with metrics that approximate them. Fourth, the QP framework and quality score do not function as black boxes and thus produce explainable scores. Consequently, the QP instantiations can be critiqued, extended, or replaced if the requirements of the user demands such. Fifth, we compared seeds from Twitter Micro-collections that were scored/selected with QPs against seeds collected by human experts and scraped from Google (also scored/selected with QPs). Micro-collections (e.g., threaded conversations from single/multiple users) are social media posts that contain URLs that are gathered by humans (as opposed to search engines) as a demonstration of domain expertise and editorial activity \cite{nwalaJCDL2019}. Our evaluation results showed that QPs resulted in the selection of quality seeds with increased precision (by $\sim$0.13) when novelty is and is not prioritized. Our code and evaluation dataset (generated between 2014 and 2020), are publicly available \cite{JCDL2021}. The dataset is comprised of 1,552 seeds from reference collections (Google and manual selection by experts) and 2,027 seeds from 4,209 tweets.
\section{Related work}
\label{sec:relatedwork}
The goal of determining the quality of URLs found in social media posts is one shared by Web archivists (Section \ref{relWorkSec:seedSelQualityAssmt}) and social media researchers (Section \ref{relWorkSec:fakeNews} and \ref{relWorkSec:ranking}).
\subsection{Seed selection and quality assessment}
\label{relWorkSec:seedSelQualityAssmt}
Risse et al. \cite{risse2014you} addressed the problem of determining attributes indicative of quality for Web archive seeds in the digital humanities domain by surveying scientists in social sciences, historical sciences, and law. Among others, they proposed that seeds should cover the evolution (topical dimension) of an event over time (temporal dimension) as opposed to a time or topic slice which gives an incomplete picture. Nwala et al. \cite{nwala2017local} proposed extracting seeds from local news sources for local events by showing that collections from local news articles produced older and lesser exposed stories than their non-local counterparts. These contributions from the Web archiving community inform the Qualities Proxies' $geographical$ and $temporal$ proximity classes.
\subsection{Content credibility and fake news detection}
\label{relWorkSec:fakeNews}
There are many studies that propose methods for assessing the credibility of information on social media platforms such as Twitter. These mostly focus on the content (e.g., text) of the social media posts and not the URLs (seeds) found in the posts, which is the focus of the QP framework. However, we posit that the quality of a seed URL can be approximated by the quality of the social media post that embeds it, and thus the following research is relevant to the QPs. Castillo et al. \cite{castillo2011information} adopted Merriam Webster's definition of credibility (``\textit{offering reasonable grounds for being believed}'') and automatically assessed the credibility of a given set of tweets and classified them as $credible$ or \textit{not credible} based on features extracted from the \textit{tweet content}, $author$, and $propagation$. Similarly, to help extract credible tweets from a flood of tweets triggered by a major news event such as a natural disaster, Gupta and Kumaraguru \cite{gupta2012credibility} identified content (\textit{frequency of unique characters}, \textit{swear words} etc.) and user-based features (e.g., \textit{number of followers}) to train a supervised machine learning and relevance feedback system. Their analysis of tweets posted about 14 high impact events in 2011 showed that on average 30\% tweets contained situational information about the event, 14\% were spam, and 17\% contained situational awareness information that was credible. 

Bozarth and Budak \cite{bozarth2020toward} demonstrated the importance of an evaluation framework of fake news detection to complement traditional evaluation metrics like F1 and precision. They used error analysis to show that classifiers' performance varied depending on multiple factors including the choice of dataset and how the training data is split (e.g., 5-fold, 80/20). Shu et al. \cite{shu2020hierarchical} approached fake news detection by studying the pattern of how news spreads on social media from news publishers to social media posts that (re)tweet content or conversation threads between accounts. Their experiments showed that the multi-level propagation network approach for fake news detection out-performed state-of-the-art fake news detection methods by at least 1.7\% with an average F1>0.84. Similarly, Bal et al. \cite{bal2020analysing} proposed an attention-based deep learning model that first identified tweets about the cause or cures of cancer and subsequently labeled those that spread misinformation.

The Quality Proxies framework includes the $reputation$ class that approximates the credibility of the domain of seed URLs in social media posts. In this research, we did not develop a method for directly assigning credibility scores to seed domains but instead approximate them (Section \ref{sec:qpRepQualProx}) by counting how often the domains were cited by Wikipedia editors. However, the user of the framework could replace our instantiation for approximating $reputation$ with a different method such as those discussed in this section.
\subsection{Ranking social media users and/or content}
\label{relWorkSec:ranking}
Popularity is widely used as a proxy for quality and credibility \cite{ciampaglia2018algorithmic}. However, algorithms that use popularity for ranking can be exploited \cite{ferrara2016rise, ratkiewicz2011detecting}. Consequently, Abbasi and Liu \cite{abbasi2013measuring} introduced the CredRank algorithm for ranking social media users by a credibility score determined by their online behavior. CredRank first attempts to detect coordinated accounts that artificially inflate the popularity of some content. Next, it suppresses the votes of the culprits in order to give preference to independently popular (credible) content. Similarly, Pal and Counts \cite{pal2011identifying} addressed the task of identifying social media users who are authorities for a given topic by proposing a set of features for characterizing social media authors, such as \textit{original tweets}, \textit{conversational tweets}, \textit{repeated tweets}, \textit{mentions}, etc. Next, using probabilistic clustering over these features, they ranked users in order to identify authorities. 

Agichtein et al. \cite{agichtein2008finding} explored using community feedback to identify high-quality content on question/answering social media platforms. They proposed a graph-based model of contributor relationships and combined it with content and usage statistics to identify quality questions and answers applied on Yahoo! Answers. Becker et al. \cite{becker2011selecting} explored centrality-based approaches (\textit{Centroid}, \textit{LexRank}, and \textit{Degree}) for identifying high quality text context from tweets for events. They defined high quality tweet text as that which contains relevant information (event time, location, participants, opinions) that is most representative of an event. Their results showed that the Centroid approach selected tweets most related to an event.

Bian et al. \cite{bian2009learning} addressed quantifying quality of content by combining content quality estimation with user reputation estimation in order to identify quality content and improved the accuracy (above state-of-the-art methods) of search over community question answering archives. Similarly, Canini et al.'s \cite{canini2011finding} investigation into the factors that affect the credibility of users on social media led to a conclusion that both the topical content of information sources and social network structure affect credibility. This conclusion led them to design a method that automatically identifies and ranks Twitter users according to their relevance and expertise for a given topic. 

Similar to the $reputation$ class, the QP framework includes the \textit{subject-expert} class that approximates the subject-expertise of the domain (e.g., \texttt{cdc.gov}) of a seed URL found in a social media posts. We did not develop a method for directly assigning subject-expertise scores to seed domains, but instead approximated them (Section \ref{sec:qpSubjQualProx}) with search engines such as Google. However, the user of the framework could replace our instantiation for approximating \textit{subject-expert} with a different method such as those discussed in this section.
\vspace{-2mm}
\section{The Quality Proxies Framework}
\label{sec:qpFramework}
The seed URL quality problem is not unique to social media. A Search Engine (SE) must return a small list of URLs (from possibly millions of candidates) to fulfill an informational request encoded in a search query. It starts by identifying relevant pages --- \textbf{$relevance$} is a proxy for quality --- but goes beyond relevance to rank webpages with a preference for popular webpages. In summary, SEs use \textbf{$popularity$} as one method to approximate quality. This is reasonable since one can argue that popularity is the reward for quality. Ciampaglia et al. \cite{ciampaglia2018algorithmic} argue that measures such as the \textit{citation rates of scientific papers}, \textit{number of downloads of a song}, or the \textit{number of social media followers} are often used in the absence of measurable notions of quality. Additionally, the goal of algorithms that favor popular items ``is to identify high-quality items such as reliable news, credible information sources, and important discoveries - in short, high-quality content should rank at the top.''

However, popularity does not always mean quality since popularity could be exploited by fake reviews, social bots, and astroturf campaigns \cite{ferrara2016rise, ratkiewicz2011detecting}. In SEs, the use of popularity in ranking algorithms was alleged to reduce the novelty, a problem that could however be mitigated by diverse user queries \cite{fortunato2006topical}. Consequently, we argue that popularity is not sufficient as a QP and explore additional non-popularity based QPs (e.g., \textit{proximity} and \textit{reputation}, Section \ref{sec:qpNonPopularityQualityProxy}). Nonetheless, popularity remains one effective proxy for quality, and as such is included in our QP framework.

In this section, we introduce the classes (e.g., $popularity$) that make up the QP framework and the metrics (e.g., $likes$ $lk$) used to instantiate them. For each seed URL, the values (all normalized between 0 and 1) for each metric are used to populate the seed QP vector which holds the multi-dimensional quality score of the seed.
\vspace{-3.3mm}
\subsection{Popularity-based Quality Proxy classes}
\label{sec:qpPopularity}
There are generally two approaches toward quantifying the popularity of URLs. The first computationally-expensive, link-based (e.g., PageRank) approach \cite{cho1998efficient} utilizes the link structure of the Web to assign weights to webpages. We adopted the second lesser computationally-expensive approach that leverages social media post statistics to assign popularity scores to URLs found in social media posts. Social media posts often keep statistics that track the number of times a post is shared (a ``retweet'' on Twitter), liked, or replied to. Transitively, the popularity of URLs from social media posts could be derived from the social media post statistics \cite{gupta2012credibility, duan2010empirical, nagmoti2010ranking} and also used to rank posts.
\vspace{-1mm}
\subsection{Post popularity Quality Proxy classes}
The post popularity classes assign popularity to a seed URL by quantifying the popularity of the post(s) containing the URL. We instantiated them with metrics that count how many people replied to ($replies$ $rp$), shared ($shares$ $sh$), and liked ($likes$ $lk$) a social media post. All of these are normalized ($x_{normalized} = \frac{x - min_X}{max_X - min_X}$) in the QP vectors for seeds.
\subsection{Author popularity $ap$ Quality Proxy}
\label{sec:authorPopularity}
The \textit{author popularity} $ap$ QP expresses the popularity of the author(s) who created the social media post(s) containing the seed URL. For example, Twitter and Instagram count $followers$ (in-degree), and $following$ or $friends$ (out-degree). Unlike Twitter, which separately counts in-degree and out-degree, Facebook only counts $friends$ (in-degree and out-degree).

For social media platforms like Facebook with bi-directional links, we instantiated $ap$ with the normalized count of $friends$. For social media platforms like Twitter, $ap = $ in-degree $-$ out-degree (e.g., $followers - following$ for Twitter) normalized. If the in-degree $ < $ out-degree, then $ap < 0$. To fix this, the offset (the absolute value of smallest difference between in-degree and out-degree) is added to each difference before normalization. Given a set of social media posts $P$, let $in_i$ and $out_i$ represent the in-degree and out-degree of social media post $i$, respectively, Eqn. \ref{eqn:ap} instantiates $ap$.

\begin{equation}
\begin{array}{l}
  \mbox{Author (}ap_i\mbox{) OR domain popularity (}dp_i) = \frac{d_i + offset}{max\underset{i \in P}{(d_i)} - min\underset{i \in P}{(d_i)} }
  \\\\
  d_i = in_i - out_i
  \\
  offset = 
  \begin{cases}
       0                    & \text{; if $min\underset{i \in P}{(d_i)} \geq 0$} \\
       |min\underset{i \in P}{(d_i)}| & \text{; otherwise} \\
  \end{cases}
  \label{eqn:ap}
\end{array}
\end{equation}

\subsection{Domain popularity $dp$ Quality Proxy}
The \textit{domain popularity} $dp$ QP quantifies the popularity of a seed's domain. We instantiated it with Eqn. \ref{eqn:ap} by approximating the popularity of the social media account (e.g., \texttt{@CDCgov}) associated with the seed domain (e.g., \texttt{cdc.gov}). To calculate $dp$ for a seed (e.g., \textcolor{blue}{\url{https://www.cdc.gov/coronavirus/2019-nCoV/}}), utilizing Twitter as example, first, we must find the social media account (\textcolor{blue}{\url{https://twitter.com/CDCgov}}) associated with the domain (e.g., \texttt{cdc.gov}). This is done by finding a bi-directional link between the social media account and the seed's website. For example, \texttt{cdc.gov} domain links to \texttt{@CDCgov} Twitter account and vice versa. Second, we extract the $in-$ and $out-$degree details from the account. Third, we apply Eqn. \ref{eqn:ap}.
\section{Non-popularity QP classes}
\label{sec:qpNonPopularityQualityProxy}
We have already discussed some limitations of popularity as a proxy for quality such as the artificial manipulation of popularity by fake reviews, social bots, and astroturf campaigns. In addition to these, it is important to note that \textbf{not all authoritative or credible sources are popular}. For example, MLive, a local media organization located in Michigan, the epicenter of the \textit{Flint Water Crisis}, is less popular than CNN, a national/international news organization, so one can argue that MLive is a local authority on topics about the \textit{Flint Water Crisis}, more so than CNN. In fact, according to Denise Robbins, it took the national media one year after the E. coli outbreak to report the Flint story \cite{DRobbins}. Consequently, it is pertinent to quantify quality (e.g. authority) across other classes in addition to popularity. This is the rationale for the following non-popularity based Quality Proxy classes (Table \ref{tab:seedQPs}, No. 4 -- 12).
\subsection{Geographical ($ge$ = $ge_a$, $ge_d$) Quality Proxy}
\label{sec:qpGeoQualProx}
Stories and events are often associated with some geographical location. For example, \textit{Hurricane Harvey} made landfall in Texas in August 2017. The \textit{geographical} $ge$ QP gives credit to a local source (local authority) when geographical location information is present. The local source could be an individual ($ge_a$ - author geographical QP) or an organization ($ge_d$ - domain geographical QP). For example, if our reference epicenter is Texas, USA, given two seeds about \textit{Hurricane Harvey} from CNN and TexasObserver (Texas local media), the $ge_d$ QP would assign a higher value to TexasObserver. Similarly, given two individuals, a resident of Rockport, Texas, and a resident of San Francisco, California, the $ge_a$ would give more credit to the Rockport resident.

We instantiated the $ge_a$ (or $ge_d$) QP with the normalized ([0, 1]) distance (measured with the Haversine formula) between a reference epicenter and the geo-location associated with the post author (for $ge_a$) or social media account associated via a bi-directional link (similar to $dp$) with the seed domain (for $ge_d$). We utilized the Google Maps Services Places API \cite{googleMapsAPI} to normalize names (e.g., ``NYC'' and ``New York'') into a single name and geo-coordinates.
\vspace{-1mm}
\subsection{Temporal $tp$ Quality Proxy}
\label{sec:qpTempQualProx}
The stories and events often happen at a place (or places), but always happen at some time. After the occurrence of the event or before its occurrence, news organizations report the story or event. For example, some of the earliest reports of the \textit{Flint Water Crisis} story are from Mlive. The temporal Quality Proxy $tp$ rewards seeds published ``early,'' when a priori information about what constitutes early is present. We instantiate it with the normalized time difference between the publication date of the seed and the reference point considered early.
\vspace{-1mm}
\subsection{Subject-expert $su$ Quality Proxy}
\label{sec:qpSubjQualProx}
The \textit{subject expert} $su$ QP approximates the subject expertise of a seed's domain. For example, given two seeds about the Coronavirus, one from the CDC and another from the blog of a high school senior, $su$ would assign the CDC a higher subject expert score since the CDC is an authority on health topics. However, how does one measure the subject expertise of \texttt{cdc.gov}? 

We instantiated $su$ based on this simple assumption: \textit{A subject expert often has more to say about their subject of expertise}. This means, if indeed the CDC is an expert on Coronavirus, we would expect to see many more reports from the CDC about Coronavirus than say ESPN. We acknowledge that this is a simplifying assumption that could be exploited. We used Document Frequency (DF) to instantiate the subject expertise $su$ of the domain of a seed. We extract DF scores by counting the number of result pages returned by Google for a given query normalized by the total number of pages indexed by the search engine for the site. This normalization is needed to avoid giving more advantage to larger websites.
\vspace{-1mm}
\subsection{Retrievability $rt$ Quality Proxy}
\label{sec:qpRetQualProx}
Seeds extracted from social media posts could also be scraped from Search Engine Result Pages (SERPs). The $retrievability$ $rt$ QP approximates how easy a seed is to find \cite{azzopardi2008retrievability}. For example, Wikipedia pages for various entities (e.g., political figures) are often placed on the front page of SERPs, meaning they have high retrievability. For this reason, $rt$ quantifies the level of difficulty of finding a seed. It is often a desirable quality to identify relevant seeds that are not easy to find to increase the novelty of a collection. We instantiated $rt$ of a seed $d$ (e.g., \textcolor{blue}{\url{https://www.cdc.gov/vhf/ebola/index.html}}) with its reciprocal rank $\frac{1}{rank_d}$ (e.g., 1/2) when searching the first $k$ (e.g., $k = 20$) Google SERPs for the seed with the query (e.g., ``ebola virus'') used to extract seeds.
\vspace{-1mm}
\subsection{Reputation ($re$ = $re_b$, $re_n$) Quality Proxy}
\label{sec:qpRepQualProx}
Social media seed URLs originate from sources with varying reputations. Given two URLs about Coronavirus, one from InfoWars (promotes conspiracy theories \cite{ramadan2016manufacturing}) and another from the CDC, it would be problematic to consider the quality of information derived from both sources equal. Similar to the \textit{subject expert} QP, the \textit{reputation} $re$ approximates the reputation of the domain of seeds. 

We defined two kinds of reputation QPs. First, \textit{reputation-broad} $re_b$ attributes reputation to the domain of a seed for having a record of publishing content about a topic (e.g., \textit{health topic}), while \textit{reputation-narrow} $re_n$ attributes reputation to the domain of a seed for having a record of publishing content focused specifically on a story (e.g., \textit{Coronavirus}). But the question remains, how does one approximate reputation? We instantiated $re$ by leveraging the expertise of Wikipedia editors. We posit that \textit{Wikipedia editors presumably sample reputable sources} \cite{chesney2006empirical}. Specifically, the reputation of the domain of a seed corresponds to the fraction of times it was cited as a reference from a gold-standard set of Wikipedia articles.

For $re_b$, the gold-standard is represented by a collection of Wikipedia articles that focus on the topic (e.g., \textit{Disease outbreaks}) of the seed. For $re_n$, the gold-standard is represented by the canonical Wikipedia page for the story. The canonical page can be found by searching for the top ranked Wikipedia page for the query (e.g., ``ebola virus outbreak'') representing the topic. To assign $re_b$ or $re_n$ to the domain of a seed, we extracted the URIs from the references of the reputation gold-standard Wikipedia articles and calculated the fraction of times each domain was referenced. For example, in our reputation gold-standard for the \textit{Disease outbreaks} (\textcolor{blue}{\url{https://en.wikipedia.org/wiki/List_of_epidemics}}) topic, \texttt{cdc.gov} appeared 42 out of 57 gold-standard articles. Therefore, the \texttt{cdc.gov} domain has a $re_b$ score of 0.74. The \texttt{cdc.gov} domain appears 14 times out of 720 references in the canonical \textit{2014 Western African Ebola Virus Outbreak} Wikipedia (\textcolor{blue}{\url{https://en.wikipedia.org/wiki/Western_African_Ebola_virus_epidemic}}) page, and thus, $re_n = 0.02$. In contrast, for \texttt{sputniknews.com}, $re_b = 0.02$ (1/57) and $re_n = 0$ (0/720). 
\vspace{-2.9mm}
\subsection{Relevance $rl$ Quality Proxy}
\label{sec:qpRelQualProx}
The $relevance$ $rl$ QP measures the degree to which a seed is on-topic. A seed that receives high marks across all the other QP vector dimensions remains non-relevant if it is off-topic. We approximate relevance by simply measuring the cosine similarity between a seed's document vector and a gold-standard document vector that captures our definition of relevance. The gold-standard is created by concatenating the text of hand-selected documents (Section \ref{sec:qualityNonNovel}, Step 1) that are relevant to a topic, and creating a feature (vocabulary) vector consisting of the TF or TFIDF weights of the terms in the concatenated document.

\subsection{Scarcity $sc$ Quality Proxy}
\label{sec:qpScaQualProx}
The $scarcity$ $sc$ QP rewards seeds from domains that are rare in a collection of seeds. It is not surprising to find multiple seeds from news organizations (e.g., \texttt{cnn.com}, \texttt{foxnews.com}, \texttt{bbc.co.uk}) for news topics. Sometimes far-reaching news events are covered by organizations for which news is not their primary domain (e.g., \texttt{eonline.com} and \texttt{espn.com}) and which may offer a novel reporting perspective. The $sc$ QP was created to surface such seeds and is approximated by $1 - \frac{|d_s|}{N}$, where $d_s$ is the frequency of a seed's domain out of $N$ total domains. 
\section{Additional $QPs$: Flipping $\overline{QPs}$}
Thus far, the Quality Proxies have been presented with the assumption that the higher the QP value, the better the trait the QP captures. For example, a high author popularity $ap$ score is a desirable trait, and a low author popularity score is not a desirable trait. However, desirability can be subjective. This means a curator might desire to surface seeds from authors that are not popular in an effort to amplify the voices of obscure users. Consequently, this requires flipping the direction of the reward system of the QP under consideration. For example, before flipping, the most popular author would have $ap = 1$, but if we flipped (represented with bar over the QP) the $ap$ Quality Proxy, $\overline{ap} = 0$ is assigned to the most popular author. Since all the quality proxies were designed to fall within [0, 1], a QP $qp$ is simply flipped by $1 - qp$; $\overline{qp} = 1 - qp$.

The ability to flip QPs provides us with additional QPs ($\overline{rp}$, $\overline{ge_a}$, $\overline{rt}$, etc). But it must be noted that the unflipped ($qp$) state and the flipped ($\overline{qp}$) state of QPs are mutually exclusive.
\section{The QP vector and scoring seeds}
\label{sec:qualityProxiesComp}
The seed Quality Proxy vector $\boldsymbol{q}$ is a 14-dimensional vector ($\boldsymbol{q} \in \mathbb{R}^{14}$) of all the values of metrics (e.g., $rp$, $sh$, $ge_a$, $re_b$) that instantiate the classes ($popularity$, $proximity$, and $uncategorized$) of the Quality Proxies framework. The QP vector of a seed assigns quality scores to a seed across multiple dimensions. Each metric's value $q_i \in \boldsymbol{q}$ expresses some quality trait of a seed and is normalized ($q_i \in [0, 1]$) such that 0 represents lowest quality and 1 represents highest quality. The dimensions of the QP vector representing multiple quality traits can be combined into a single score $q_f$ that can be used to score and/or rank seeds. We instantiated the QP score function $q_f$ (Eqn. \ref{eqn:qpScore}) of a seed simply with the 2-norm of the n-dimensional QP vector $\boldsymbol{q}$ of the seed.
\begin{equation}
  q_f = |\boldsymbol{q}|_2 = \sqrt{\sum_{i=1}^{n} q_i^2}
  \label{eqn:qpScore}
\end{equation}
A user can control the relative importance of the metrics of $\boldsymbol{q}$ depending on prior information or specific needs. Therefore, one can multiply a weight vector $\boldsymbol{w} \in \mathbb{R}^{14}$ ($\sum_{i=1}^{n} w_i = 1$) with $\boldsymbol{q}$ ($q'_i = q_iw_i$) to reflect the importance of each metric to obtain a new Quality Proxy scores $q'_f$. The weight vector can also be used to switch off specific metrics. For example to switch off $q_i$, we set $w_i = 0$, such that $q_iw_i = 0$.

\begin{table}
\setlength{\tabcolsep}{1pt}
\centering
\caption{\textcolor{red}{\textit{2020 Coronavirus Pandemic}}: top five seeds extracted by combining three popularity-based QPs $\{rp$, $sh$, $lk\}$ to produce a single quality score ($q_f$ - Eqn. \ref{eqn:qpScore}) and ranking the seeds by their scores. Popularity QPs unsurprisingly give more credit to seeds from popular (well-known) domains/users.}
\begin{tabular}{|c|c|c|c|c|c.c|c|c|}
\hline
\multirow{2}{*}{\textbf{\#}} & \multirow{2}{*}{ \makecell{\textbf{\textcolor{blue}{domain}:} title\\(\textcolor{red}{\texttt{user's twitter handle}}) }} & \multicolumn{4}{c.}{ \makecell{\textbf{QP score $q_f$}\\\textbf{\& QP values}} }           & \multicolumn{3}{c|}{ \makecell{\textbf{QP values}\\\textbf{(thousands)}} }                  \\ \cline{3-9} 
                             &                                & $q_f$ & \textbf{rp} & \textbf{sh} & \textbf{lk} & \textbf{rp} & \textbf{sh} & \textbf{lk} \\ \hline                                                                                                                                                   %  nm & rp  & sh   & lk  & rp     & sh    & lk
1                            & \href{https://twitter.com/HillaryClinton/status/1248270935846723591}{\makecell{\textcolor{blue}{\textbf{reuters.com}}: Most Americans,\\unlike Trump, want mail-in...\\(\textcolor{red}{\texttt{@HillaryClinton}})}}                             & 1.0 & 1.0 & 1.0 & 1.0 & 13.4  & 31.7 & 101 \\ \hline
2                            & \href{https://twitter.com/CNBC/status/1248303279169245184}{\makecell{\textcolor{blue}{\textbf{cnbc.com}}: Chamath Palihapitiya:\\US shouldn't bail out\\hedge funds, billionaires (\textcolor{red}{\texttt{@CNBC}})}}                            & .54 & .26 & .67 & .59 & 3.48  & 21.3 & 59.9 \\ \hline
3                            & \href{https://twitter.com/nicktolhurst/status/1248294987529367553}{\makecell{\textcolor{blue}{\textbf{gov.uk}}: New immigration system:\\ what you need to know\\(\textcolor{red}{\texttt{@nicktolhurst}})}}                                     & .39 & .56 & .23 & .30 & 7.54  & 7.20 & 30.1 \\ \hline
4                            & \href{https://twitter.com/SenSanders/status/1248318315489001472}{\makecell{\textcolor{blue}{\textbf{washingtonpost.com}}:\\When coronavirus hits, but the\\water is shut off (\textcolor{red}{\texttt{@SenSanders}})}}                           & .32 & .08 & .32 & .46 & 1.09  & 9.99 & 46.4 \\ \hline
5                            & \href{https://twitter.com/TheRickWilson/status/1248238176122044416}{ \makecell{\textcolor{blue}{\textbf{wsj.com}}: Trump's Wasted\\Briefings (\textcolor{red}{\texttt{@TheRickWilson}})} }                                                       & .25 & .15 & .28 & .29 & 2.07  & 8.97 & 29.8 \\ \hline
\end{tabular}
\label{tab:qpStudyPopularity}
\end{table}

%y: \multicolumn{3}{c|}{ \textbf{QP Normalized} }
\begin{table}
\small
\setlength{\tabcolsep}{1pt}
\centering
\caption{\textcolor{red}{\textit{2020 Coronavirus Pandemic}}: top five seeds extracted by combining geographical QPs $\overline{ge_a}$ and $\overline{ge_d}$ (seeds from users and domains distant from New York) and ranking the seeds by their QP scores ($q_f$ - Eqn. \ref{eqn:qpScore}). The table illustrates what seeds users from distant geographical regions share.}
\begin{tabular}{|c|c|c|c|c|}
\hline
\textbf{\#} & \makecell{\textbf{\textcolor{blue}{domain (domain org. location)}}: title\\(\textcolor{red}{\texttt{user's twitter handle}, user's location}) }  & \multicolumn{3}{c|}{ }                         \\ \hline
\multicolumn{2}{|c|}{ \textbf{Epicenter: New York City} } & $q_f$ & $\boldsymbol{\overline{ge_a}}$ & $\boldsymbol{\overline{ge_d}}$ \\ \hline

1                            & \href{https://twitter.com/icklato/status/1248246331111440384}{ \makecell{\textcolor{blue}{\textbf{thejakartapost.com (Jakarta)}}: Finland\\discovers masks bought from China not\\hospital-safe (\textcolor{red}{\texttt{@ick\_forPH}, Philippines}) }}                                                                                                                                                    & 0.92 & 0.83 & 1.00 \\ \hline
2                            & \href{https://twitter.com/rapplerdotcom/status/1248238352597426178}{ \makecell{\textcolor{blue}{\textbf{rappler.com (Philippines)}}: FACT CHECK: Duque\\claims PH has `low' coronavirus infection\\(\textcolor{red}{\texttt{@rapplerdotcom}, Philippines})}}                                                                                           & 0.84 & 0.83 & 0.86 \\ \hline
3                            & \href{https://twitter.com/GHNeale/status/1248310339822850053}{ \makecell{\textcolor{blue}{\textbf{bylinetimes.com (London)}}: COVID-19 SPECIAL\\INVESTIGATION: Leaked Home Office\\...(\textcolor{red}{\texttt{@GHNeale}, NA})}}                                    & 0.75 & 1.00 & 0.34 \\ \hline
4                            & \href{https://twitter.com/OfficialKRU/status/1248251864258359296}{ \makecell{\textcolor{blue}{\textbf{kru.co.ke (Nairobi)}}: Kenya Rugby Union\\announces cancellation of 2019/20\\season as Corona virus...(\textcolor{red}{\texttt{@OfficialKRU}, Nairobi})}}                                                                                                           & 0.72 & 0.71 & 0.73 \\ \hline
5                            & \href{https://twitter.com/_lameckongeri/status/1248291220427866119}{ \makecell{\textcolor{blue}{\textbf{jesusislordradio.info (Nakuru, Kenya)}}:\\Welcome To Jesus Is Lord Radio\\(\textcolor{red}{\texttt{@\_lameckongeri}, Kisii, Kenya})} }                                                                                                                                                                                                                                                    & 0.71 & 0.69 & 0.72 \\ \hline

\end{tabular}
\label{tab:qpStudyGeo}
\end{table}

\begin{table}
\setlength{\tabcolsep}{1.5pt}
\centering
\caption{\textcolor{red}{\textit{2020 Coronavirus Pandemic}}: top five seeds with the highest \textit{broad reputation} $re_b$ values. For a single seed (e.g., \textcolor{blue}{\url{https://www.ncbi.nlm.nih.gov/pmc/articles/PMC1592694/}} from \#2), the $re_b$ score (e.g., 0.81) was approximated by counting the number of times (\textbf{Hits}) the seed domain (e.g., \texttt{nih.gov}) was cited (e.g., 46 times) in a reputation gold standard of 57 representative Wikipedia documents (one vote per document) about \textit{Disease outbreaks}.}
\begin{tabular}{|c|c|c|c|}
\hline
\textbf{\#} &  \textbf{\textcolor{blue}{domain}}: title (\textcolor{red}{user's twitter handle})  & $\boldsymbol{re_b}$ & \textbf{Hits}                  \\ \hline
                                                                                                                                              %  nm & rp  & sh   & lk  & rp     & sh    & lk
1                            & \href{https://twitter.com/SergioBowers1/status/1248267420935974913}{ \makecell{\textcolor{blue}{\textbf{who.int}}: Tobacco\\(\textcolor{red}{@SergioBowers1})}}                             & 0.82 & 47 \\ \hline
2                            & \href{https://twitter.com/HITNTNotTalkin/status/1248315500913987590}{ \makecell{\textcolor{blue}{\textbf{nih.gov}}: Ventilator-Associated Pneumonia:\\Diagnosis, Treatment, and Prevention\\(\textcolor{red}{@HITNTNotTalkin})}}                            & 0.81 & 46 \\ \hline
3                            & \href{https://twitter.com/2020DoOver/status/1248290612698320896}{ \makecell{\textcolor{blue}{\textbf{cdc.gov:}}: 2009 H1N1 Pandemic\\(H1N1pdm09 virus) Pandemic\\Influenza (Flu) (\textcolor{red}{@2020DoOver})}}                                & 0.74 & 42 \\ \hline
4                            & \href{https://twitter.com/peabodypress/status/1248320850903826433}{ \makecell{\textcolor{blue}{\textbf{cdc.gov}}: Legal Authorities for\\Isolation and Quarantine (\textcolor{red}{@peabodypress})}}                              & 0.74 & 42 \\ \hline
5                            & \href{https://twitter.com/Rick51224214/status/1248308682003222529}{ \makecell{\textcolor{blue}{\textbf{cdc.gov}}: 2019-2020 U.S. Flu Season:\\Preliminary Burden Estimates\\(\textcolor{red}{@Rick51224214})} }                             & 0.74 & 42 \\ \hline
\end{tabular}
\label{tab:qpStudyRe}
\end{table}

\begin{table}
\small
\setlength{\tabcolsep}{1.5pt}
\centering
\caption{\textcolor{red}{\textit{Flint Water Crisis}}: top five seeds extracted by combining relevance $rl$ and geographical $ge_d$ QPs to surface local media (e.g., \texttt{detroitnews.com}) in Flint, Michigan.}
\begin{tabular}{|c|c|c|c|c|}
\hline
\textbf{\#} & \makecell{\textbf{\textcolor{blue}{domain (domain org. location)}}: title\\(\textcolor{red}{\texttt{user's twitter handle}, user's location}) } & $q_f$ & \textbf{rl} & $\boldsymbol{ge_d}$  \\ \hline
%\multicolumn{2}{|c|}{ \textbf{Section I} } & $q_f$ & \textbf{rl} & $\boldsymbol{ge_d}$  \\ \hline                                                                                                                                                   %  nm & rp  & sh   & lk  & rp     & sh    & lk
1                            & \href{https://twitter.com/PhilRevard/status/1056920816892461058}{ \makecell{\textcolor{blue}{\textbf{mlive.com (Michigan)}}: As Flint was slowly\\poisoned, Snyder's inner circle\\failed to act (\textcolor{red}{\texttt{@PhilRevard}, Michigan})}}                                                                              & 0.91 & 0.85 & 0.97 \\ \hline
2                            & \href{https://twitter.com/LOLGOP/status/1163092107348889600}{ \makecell{\textcolor{blue}{\textbf{eclectablog.com (Ann Arbor, Michigan)}}:\\The deceptive corporatist rewriting of the\\history of the \#FlintWaterCrisis is in full swing\\(\textcolor{red}{\texttt{@LOLGOP}, Ann Arbor, Michigan})}}                                                                          & 0.86 & 0.72 & 0.99 \\ \hline
3                            & \href{https://twitter.com/PhilRevard/status/1053332899603365888}{ \makecell{\textcolor{blue}{\textbf{detroitnews.com (Detroit, Michigan)}}: AG's office\\got Flint complaints a year before\\probe (\textcolor{red}{\texttt{@PhilRevard}, Michigan})}}                                                                             & 0.85 & 0.68 & 0.99 \\ \hline
4                            & \href{https://twitter.com/jmlarkin/status/1145806004027682816}{ \makecell{\textcolor{blue}{\textbf{michiganadvance.com (Michigan)}}: Judge allows\\Flint water class-action lawsuit to\\proceed, adds Snyder...(\textcolor{red}{\texttt{@jmlarkin}, Cambridge, MA})}}                                                                            & 0.84 & 0.70 & 0.97 \\ \hline
5                            & \href{https://twitter.com/nreza21/status/1197722175421210624}{ \makecell{\textcolor{blue}{\textbf{michigan.gov (Michigan)}}: EGLE - Flint's water\\remains stable, continues to meet federal\\and new stricter state standards (\textcolor{red}{\texttt{@nreza21}, NA})} }                                                                                    & 0.84 & 0.69 & 0.97 \\ \hline
\end{tabular}
\label{tab:qpStudyFlint0}
\end{table}

\begin{table}
%%rl + sc
\small
\setlength{\tabcolsep}{1pt}
\centering
\caption{\textcolor{red}{\textit{Hurricane Harvey}}: top five seeds extracted by combining relevance $rl$ and the scarcity $sc$ QP, used to increase the diversity of news sources (e.g., \texttt{texasmonthly.com}, \texttt{eonline.com}, and \texttt{espn.com}) by extracting seeds from domains with the smallest representation (\textbf{Hits}) in the collection.}
\begin{tabular}{|c|c|c|c|c|c|}
\hline
\textbf{\#} & \textbf{\textcolor{blue}{domain}}: title (\textcolor{red}{user's twitter handle}) & $q_f$ & $\boldsymbol{rl}$ & $\boldsymbol{sc}$ & \textbf{Hits}                  \\ \hline
1                            & \href{https://twitter.com/TexasMonthly/status/912025515430686721}{ \makecell{\textcolor{blue}{\textbf{texasmonthly.com}}:\\Voices from the Storm (\textcolor{red}{\texttt{@TexasMonthly}})}}                                                                              & 0.71 & 0.13 & 0.99 & 1 \\ \hline
2                            & \href{https://twitter.com/texasdemocrats/status/913416109575065601}{ \makecell{\textcolor{blue}{\textbf{texasobserver.org}}: Even Hurricane Harvey\\Can't Temper GOP Hostility Toward Texas'\\Big Cities (\textcolor{red}{\texttt{@texasdemocrats}})}}                                                                      & 0.70 & 0.11 & 0.99 & 1 \\ \hline
3                            & \href{https://twitter.com/enews/status/904016734503436289}{ \makecell{\textcolor{blue}{\textbf{eonline.com}}: Taylor Swift Makes \\``Very Sizable Donation'' to Houston\\Food Bank After Hurricane Harvey (\textcolor{red}{\texttt{@enews}})}}                                                                                   & 0.70 & 0.10 & 0.99 & 1 \\ \hline
4                            & \href{https://twitter.com/SportsCenter/status/908890745905532928}{ \makecell{\textcolor{blue}{\textbf{espn.com}}: J.J. Watt's Hurricane Harvey\\charity fundraising closes with \$37M-plus\\in donations (\textcolor{red}{\texttt{@SportsCenter}})}}                                                                                   & 0.70 & 0.08 & 0.99 & 1 \\ \hline
5                            & \href{https://twitter.com/RollingStone/status/912545991311728640}{ \makecell{\textcolor{blue}{\textbf{rollingstone.com}}: Houston Astros\\After Hurricane Harvey (\textcolor{red}{\texttt{@RollingStone}})} }                                                                                      & 0.70 & 0.08 & 0.99 & 1 \\ \hline
\end{tabular}
\label{tab:qpStudyHarvey}
\end{table}

\section{Selecting seeds with QP scores}
In this section, we explore how different combinations of QPs map to different notions of quality and policies for selecting seeds for the \textit{2020 Coronavirus Pandemic} (Tables \ref{tab:qpStudyPopularity}, \ref{tab:qpStudyGeo}, \& \ref{tab:qpStudyRe}), the \textit{Flint Water Crisis} (Table \ref{tab:qpStudyFlint0}), and \textit{Hurricane Harvey (Table \ref{tab:qpStudyHarvey})}. The seed URL titles can be clicked.

Table \ref{tab:qpStudyPopularity} illustrates that a combination of popularity-based Quality Proxies $rp$, $sh$, and $lk$ unsurprisingly gives more credit to seeds from popular (well-known) domains (e.g., \texttt{reuters.com}, \texttt{cnbc.com}, \texttt{washingtonpost.com}) posted by popular authors (e.g., \texttt{@HillaryClinton}, \texttt{@CNBC}, and \texttt{@SenSanders}). Seeds from well-known domains are more likely to be replied to ($rp$), shared ($sh$), or liked ($lk$) as a result of the large audience they enjoy. Sampling seeds from popular sources could help reduce spam or reduce the number of non-credible sources.

Unlike Table \ref{tab:qpStudyPopularity}, Table \ref{tab:qpStudyGeo} shifts the reward system by prioritizing authors ($\overline{ge_a}$) and domains ($\overline{ge_d}$) geographical distant from New York. This resulted in the surfacing of authors and domains outside the United States with an international perspective. The top five authors are residents of two different countries (e.g., \texttt{@ick\_forPH} - Philippines and \texttt{@OfficialKRU} - Kenya) while the organization of the domains are from four different countries (\texttt{thejakartapost.com} - Indonesia, \texttt{rappler.com} - Philippines, \texttt{bylinetimes.com} - England, and \texttt{kru.co.ke}, \texttt{jesusislordradio.info} - Kenya).

Given the concerns of the spread of (mis/dis)information surrounding the coronavirus pandemic, curators could potentially impose stringent rules that restrict the sources of seeds to reputable sources. This \textit{reputable sources only} selection criteria aligns with the goal of the \textit{reputation-broad} QP ($re_b$). Table \ref{tab:qpStudyRe} outlines the top five seeds when seeds are scored by their respective reputation scores. For a single seed (e.g., \textcolor{blue}{\url{https://www.ncbi.nlm.nih.gov/pmc/articles/PMC1592694/}}) in Table \ref{tab:qpStudyRe}, the $re_b$ score (e.g., 0.81) was approximated by counting the number of times the seed domain (e.g., \texttt{nih.gov}) was cited (e.g., 46 times) in a reputation gold standard of 57 representative Wikipedia documents (one vote per document) about \textit{Disease outbreaks}. Accordingly, the most dominant seeds were from world-renowned health institutions such as the World Health Organization (\texttt{who.int}) which was referenced 47 times, National Institute of Health (\texttt{nih.gov}) referenced 46 times, and Centers for Disease Control and Prevention (\texttt{cdc.gov}) referenced 42 times out of 57 representative Wikipedia documents about public disease outbreaks.

Table \ref{tab:qpStudyFlint0} illustrates how the $ge_d$ QP helps surface local news organizations, such as \texttt{mlive.com}, which was critical to the coverage of the \textit{Flint Water Crisis}, by giving credit to seed domains from organizations near a geographical reference (e.g., Flint, Michigan).

Table \ref{tab:qpStudyHarvey} illustrates how the $scarcity$ $sc$ QP can help increase the diversity of sources by surfacing seeds from non-conventional news media outlets such as \textit{Taylor Swift Makes ``Very Sizable Donation'' to Houston Food Bank After Hurricane Harvey} - \texttt{eonline.com} and \textit{J.J. Watt's Hurricane Harvey charity fundraising closes with \$37M-plus in donations} - \texttt{espn.com}.

\section{Framework Evaluation}
\label{sec:frameworkEval}
The goal of this evaluation was two-fold. First, to assess the precision of the seeds selected by their Quality Proxy-assigned scores when novelty is not prioritized (Section \ref{sec:qualityNonNovel}). For brevity, we define \textit{QP seeds} as the top ranked seeds selected when seed URLs extracted from social media posts (e.g., tweets) are ranked by their QP scores. It would be unreasonable to collect QP seeds if they are of poor quality compared to expert-generated seeds. We modeled good quality with prototypical seeds referred to as $reference$ seeds scraped from Google and/or hand-selected by human-experts on Archive-It. 

Second, to assess the precision of seeds when novelty is prioritized (Section \ref{sec:qualityNovel}). It is a positive trait for QP seeds to be highly similar (low novelty) with respect to Google and/or expert-generated seeds, since this could be indicative of their high-quality. The goal of the first evaluation was to quantify the degree of similarity between QP seeds and reference seeds. However, we often need our seeds to be novel or, in other words, different from seeds produced by Google and/or experts but not at the expense of quality. Therefore, we assessed the precision of QP seeds when novelty is prioritized. Novelty of seeds was measured (Section \ref{sec:qualityNonNovel}, Step 4) by comparing them with reference (Google or Expert) seeds.

\subsection{Evaluation Dataset}
To evaluate social media seed URLs selected with their QP scores (QP seeds), we generated a dataset (Table  \ref{tab:framworkEvalDS}, \cite{JCDL2021}) consisting of seeds extracted from reference collections (Section \ref{sec:genRefCol}) and Twitter Micro-collections (Section \ref{sec:mcCol}) for multiple topics.
\subsubsection{\textbf{Generating reference (Google/Expert) seeds}}
\label{sec:genRefCol}
\textcolor{white}{.}\\
The reference collections served as baselines for defining quality. Seeds from Google were scraped, while seeds from expert-generated collections were extracted from the Archive-It API \cite{archiveItAPI}.
\subsubsection{\textbf{Extracting seeds from Micro-collections}}
\textcolor{white}{.}\\
\label{sec:mcCol}
In addition to reference Google/Expert seeds, we extracted seeds from Twitter Micro-collections to be compared to the reference seeds. Micro-collections are social media posts that contain URLs that are gathered by humans as a demonstration of domain expertise and editorial activity \cite{nwalaJCDL2019}. On Twitter, they manifest as the threaded conversations created by single or multiple users. Seeds extracted from Twitter Micro-collections are different from those scraped exclusively from SERPs \cite{NwalaJCDL2018}.

In total, the evaluation dataset (extracted from 2014 -- 2020) consisted of 1,552 seeds from reference collections, and 2,027 seeds from 4,209 tweets from Twitter Micro-collections. Even though we utilized Twitter for evaluation, our framework is applicable to other social media platforms such as Reddit and Facebook.
\begin{table}[]
\centering
\setlength{\tabcolsep}{0.8pt}
\caption{Framework evaluation dataset \cite{JCDL2021} consisting of 1,552 seeds from Reference
(Google \& Expert) collections, and 2,027 seeds from 4,209 tweets from Twitter (Top/Latest) extracted at different date ranges.}
\begin{tabular}{|c|c|c|}
\hline
\multicolumn{1}{|c|}{\textbf{Topic}} & \textbf{Extraction-Range} & \textbf{Seeds Count} \\ \hline
\multicolumn{3}{|c|}{\textbf{Reference Google Collections (808 Seeds)}}                                                                                                                      \\ \hline
  hurricane harvey                        & 2020-04-11                                                           & 199 (Page 1 - 20)     \\ \hline
  flint water crisis                     & 2020-04-10                                                           & 173 (Page 1 - 20)     \\ \hline
  coronavirus                             & 2020-04-09                                                           & 176 (Page 1 - 20)   \\ \hline
  2018 world cup                         & 2019-01-09                                                           & 112 (Page 1 - 10)   \\ \hline
  ebola virus                            & 2017-11-29                                                           & 97  (Page 1 - 10)    \\ \hline
  hurricane harvey                        & 2017-09-(02 to 29)                                                      & 51  (Page 1)          \\ \hline
\multicolumn{3}{|c|}{\textbf{Reference Expert Collection from Archive-It (744 Seeds)}}                                                                                                       \\ \hline
  coronavirus    \cite{fEvalNLM}          & 2020-03-15                                                           & 574                  \\ \hline
  hurricane harvey     \cite{fEvalVTech}  & 2017-(08-25 to 09-29)                                             & 37                   \\ \hline
  ebola virus     \cite{fEvalEbola}       & 2014-10-01                                                           & 133                  \\ \hline
\multicolumn{3}{|c|}{\textbf{Twitter-Top (1,310 Seeds, 2,221 tweets)}}                                                                                                              \\ \hline
  hurricane harvey                         & 2020-04-11                                                           & 201 (500 tweets)     \\ \hline
  flint water crisis                     & 2020-04-09                                                           & 312 (500 tweets)     \\ \hline
  coronavirus                            & 2020-04-09                                                           & 533 (500 tweets)     \\ \hline
  2018 world cup                         & 2019-01-09                                                           & 121 (500 tweets)     \\ \hline
  ebola virus                            & 2017-(11-30 to 12-31)                                             & 48  (68  tweets)     \\ \hline
  hurricane harvey                        & 2017-09-(02 to 31)                                             & 95  (153 tweets)     \\ \hline
\multicolumn{3}{|c|}{\textbf{Twitter-Latest (717 Seeds, 1,988 tweets)}}                                                                                                             \\ \hline
  flint water crisis                     & 2020-04-09                                                           & 92  (500 tweets)      \\ \hline
  coronavirus                            & 2020-04-09                                                           & 541 (500 tweets)      \\ \hline
  2018 world cup                         & 2019-01-09                                                           & 84  (488 tweets)      \\ \hline
\end{tabular}
\label{tab:framworkEvalDS}
\end{table}

\subsection{Precision when Novelty \textcolor{red}{is not} Prioritized}
\label{sec:qualityNonNovel}

The following five steps describe how we assessed the precision of QP seeds when novelty is not prioritized. 
\subsubsection*{\textbf{Step 1: Extracting Quality Proxies for Seeds}}
\textcolor{white}{.}\\
We instantiated the QP vectors for all seeds in the evaluation dataset by extracting all values for QP metrics (Table \ref{tab:seedQPs}) except \textit{subject-expert} $su$ and \textit{temporal} ($tp$) resulting in the use of 12 QPs. The $su$ instantiation with the document frequency from Google was not determined to be a dependable approximation of $su$ since it fluctuated (for the same seed) with a high variance, hence we excluded it from our evaluation. Additionally, we did not impose a temporal bias to favor old or new documents, hence we excluded $tp$.

We approximated the $relevance$ $rl$ QP with the cosine similarity between document vectors for a seed and a gold standard document created from the text of the references of Wikipedia articles corresponding to each dataset topic. The \textit{author-popularity} $ap$ QP corresponds to the popularity of the social media author of the post. Since seeds from Google and Archive-It are not posted by social media authors, we approximated the $ap$ QP with the reciprocal rank ($\frac{1}{rank_i}$) of their seeds to ensure they are comparable to QP seeds.
\begin{table}[]
\centering
\setlength{\tabcolsep}{1.5pt}
\caption{A sample of 12 QP combinatorial states for 1-combination, 2-combination, and 3-combinations. A single 1-combination or 2-combination or $r$-combination of QPs can be used to score (Eqn. \ref{eqn:qpScore}) a seed.}
\begin{tabular}{|c|c|c|c|}
\hline
\multicolumn{1}{|c|}{\textbf{\#}} & \textbf{1 - Combination} & \textbf{2 - Combinations} & \textbf{3 - Combinations} \\ \hline
1                                 & $rp$                     & $rp, lk$                  & $rp, lk, sh$              \\ \hline
2                                 & $\overline{rp}$          & $rp, sh$                  & $rp, lk, ap$              \\ \hline
3                                 & $sh$                     & $rp, ap$                  & $rp, lk, dp$              \\ \hline
4                                 & $\overline{sh}$          & $rp, dp$                  & $rp, lk, ge_a$            \\ \hline
5                                 & $lk$                     & $rp, ge_a$                & $rp, lk, ge_d$            \\ \hline
6                                 & $\overline{lk}$          & $rp, ge_d$                & $rp, sh, ap$              \\ \hline
7                                 & $ap$                     & $lk, sh$                  & $rp, sh, dp$              \\ \hline
8                                 & $\overline{ap}$          & $lk, ap$                  & $rp, sh, ge_a$            \\ \hline
9                                 & $dp$                     & $lk, dp$                  & $rp, sh, ge_d$            \\ \hline
10                                & $\overline{dp}$          & $lk, ge_a$                & $rp, ap, dp$              \\ \hline
11                                & $ge_a$                   & $lk, ge_d$                & $rp, ap, ge_a$            \\ \hline
12                                & $\overline{ge_a}$        & $sh, ap$                  & $rp, ap, ge_d$            \\ \hline
\end{tabular}
\label{qp:statesComb}
\end{table}
\subsubsection*{\textbf{Step 2: Generating QPs Combinatorial States}}
\textcolor{white}{.}\\
We utilized the 12 QPs from from the previous step to score (Eqn. \ref{eqn:qpScore}) seeds, selected the top K seeds, and compared them with top reference seeds scored with the same QPs. We did not assign weights to the Quality Proxies. Additionally, we expanded the options for scoring seeds beyond 12 QPs as follows. First, we permitted flipping the QPs, resulting in 12 additional QPs (24 QPs total). Second, we permitted using a subset of the 24 QPs, leading to a combinatorial explosion of possible QP states for scoring seeds. However, we restricted our scoring to 1-, 2-, and 3-combinations which produced a total of 2,049 possible QP combinations (e.g., $rp$, $\{\overline{ap}, ge_a\}$, $\{re_b, sc, \overline{rt}\}$) to score seeds. Table \ref{qp:statesComb} shows 72 of these combinations.

\subsubsection*{\textbf{Step 3: Scoring Seeds with a Combination of QPs}}
\textcolor{white}{.}\\
To score seeds from Twitter or reference Google or Expert collections, we first selected a single combination of Quality Proxies, for example, $\{rp, sh, lk\}$. Next, using only the QPs selected, we assigned a score to the seed with Eqn. \ref{eqn:qpScore}.

\subsubsection*{\textbf{Step 4: Twitter vs Google/Expert: comparing top K QP seeds}}
Recall the QP seeds definition: \textit{the top ranked seeds selected when seed URLs extracted from social media posts (e.g., tweets) are ranked by their QP scores}. The top K QP seeds with scores assigned by a given combination of QPs were compared to the top K reference (Google/Expert) seeds scored with the same QP combination. Comparison was done by measuring the domain (e.g., \texttt{cdc.gov}) overlap ($\frac{|A \cap B|}{min(|A|, |B|)}$) between the Twitter QP seeds and reference (Google and/or expert) seeds. We also measured the precision of the selected QP seeds and reference seeds. For precision evaluation, if the cosine similarity between a seed and the gold standard document vector is at least a predefined relevance threshold (set at 0.20 for all except \textit{Hurricane Harvey}: 0.10), the seed is considered relevant. The threshold was estimated by finding the median similarity between each gold standard document and the rest of the gold standard documents. Median scores exceeding 0.20 --- which was empirically determined to produce satisfactory baseline relevance --- were set to 0.20.

\subsubsection*{\textbf{Step 5: Seed Precision when Novelty is not Prioritized}}
\textcolor{white}{.}\\
The final process of assessing the precision of seeds when novelty is not prioritized involved reporting the average overlap and average precision for QP combinations used to score and select top K ($10 \leq K \leq 100$) seeds. This was achieved by reporting the top 10 (out of 2,049 QP combinations) overlap scores between Twitter and reference seeds and reporting Precision at K (P@K) for the associated QP combination used to score the seeds. Selecting the top 10 overlap enables us learn the precision of seeds when overlap is at its best, albeit at the expense of novelty since the higher the overlap between Twitter and reference seeds, the lower the novelty. Section \ref{sec:resultsQualityNovel} presents and discusses the results.

\subsection{Precision when Novelty \textcolor{blue}{is} Prioritized}
\label{sec:qualityNovel}
Since we consider reference seeds to be quality seeds, a high overlap between reference and Twitter QP seeds could result in a high precision of the Twitter QP seeds. However, since novelty (low overlap) is also a desirable quality of seeds, it is crucial to additionally assess the precision of Twitter QP seeds when novelty is prioritized.

The steps for assessing the precision of seeds when novelty is prioritized are the same as the previous section (when novelty is not prioritized) except for \textit{Step 5}. Instead of reporting the P@K for the associated QP combinations with the top 10 overlap scores, to prioritize novelty, we measured and reported the precision of QP combinations that produced no overlap (highest novelty) between Twitter and reference QP seeds. Section \ref{sec:resultsQualityNonNovel} discusses the results.

\section{Evaluation results and Discussion}
\label{sec:frameworkResult}
\begin{figure}
    \centering
    \includegraphics[height=0.25\textwidth]{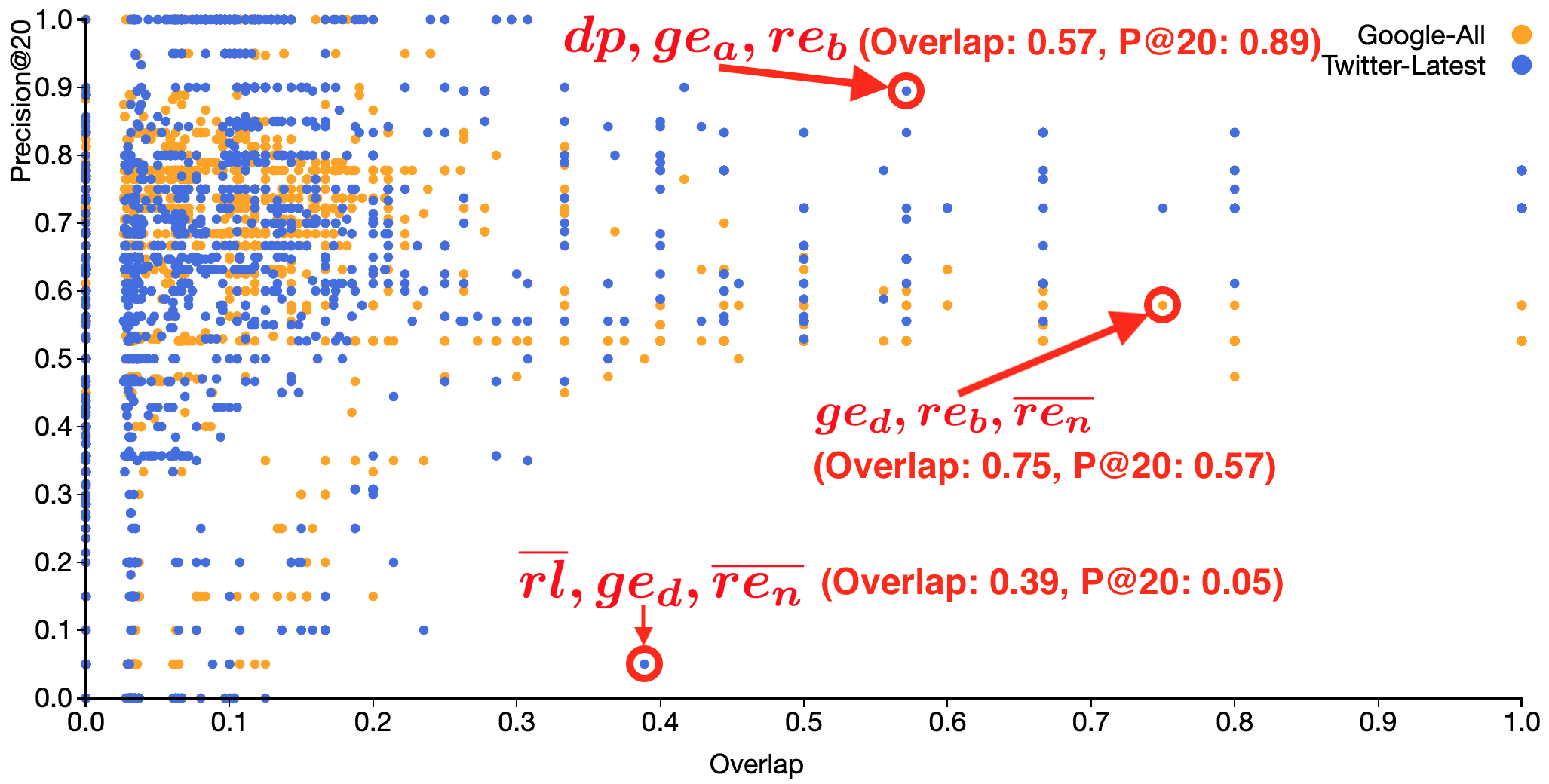}%0.32
   
    \caption{Overlap vs P@20 for Google (orange dots) and Twitter (blue dots) \textit{2020 Coronavirus Pandemic} Twitter-Latest seeds scored with different QPs. A single dot represents the overlap (X-axis) and P@20 (Y-axis) for seeds scored by a single Quality Proxy. The scatterplot shows how different combinations of QPs result in high (e.g., \textcolor{red}{$\boldsymbol{dp, ge_a, re_b}$} and \textcolor{red}{$\boldsymbol{ge_d, re_b, \overline{re_n}}$}) or low (\textcolor{red}{ $\boldsymbol{\overline{rl}, ge_d, \overline{re_n}}$ }) overlap/P@20. Unsurprisingly, \textcolor{red}{ $\boldsymbol{\overline{rl}, ge_d, \overline{re_n}}$ } resulted in a low P@20 because the $relevance$ $rl$ QP was flipped, meaning relevance was penalized.}%
    \label{fig:ovPrecScatt0}%
   
\end{figure}

Our overlap and precision results were proven to be statistically significant by a one-tailed Student's t-test with $\alpha = 0.05$ and K = 30 across all dataset topics.
\begin{comment}
non-novel
co: rl, re_b * re_n
wc: ge_a, re_b, re_n!
hh: rt, re_b
h2: rt, re_b * sc!
fl: rt, re_b
eb: rt, re_n! * sc!

novel:
co: $rl, \overline{ap}$     * $rl, \overline{rp}$
wc: $rl, \overline{sh}$
hh: $rl, \overline{rt}$
h2: $\overline{ap}, \overline{re_b}, \overline{re_n}$ * $rl, \overline{rt}$
fl: $rl, \overline{rp}, \overline{ap}$
eb: $rl, \overline{rt}$ * $rl, \overline{re_n}$
\end{comment}
\begin{figure*}
  \includegraphics[width=0.98\textwidth]{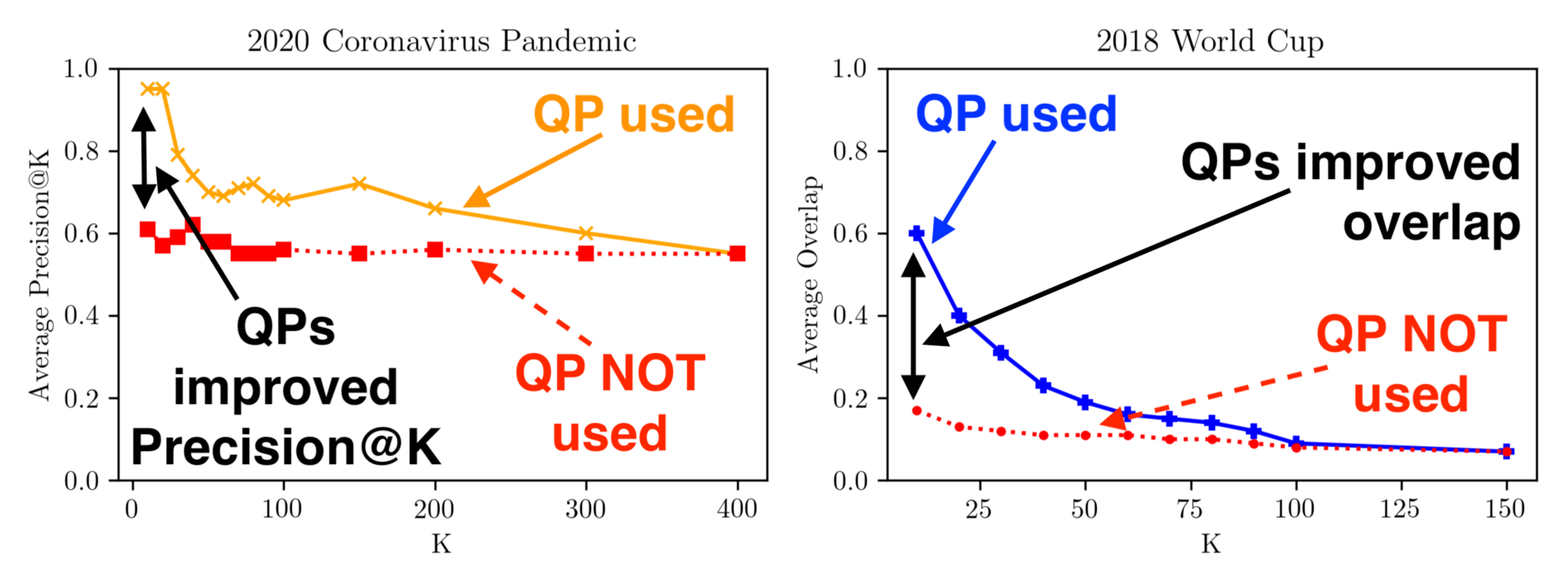} %
  \vspace*{-4mm}
  \caption{Average P@K (left) and Average overlap (with Expert seeds - right) showing that Twitter seed URLs scored and selected with Quality Proxies, improved (higher solid lines) the P@K and overlap above the baseline (lower red dotted lines) precision and overlap which did not use QPs. However, the improvement diminished as K (number of seeds) increased.}
  \label{fig:qpImprovePrecOveralap}
 
\end{figure*}

\begin{figure}
  \includegraphics[width=0.44\textwidth]{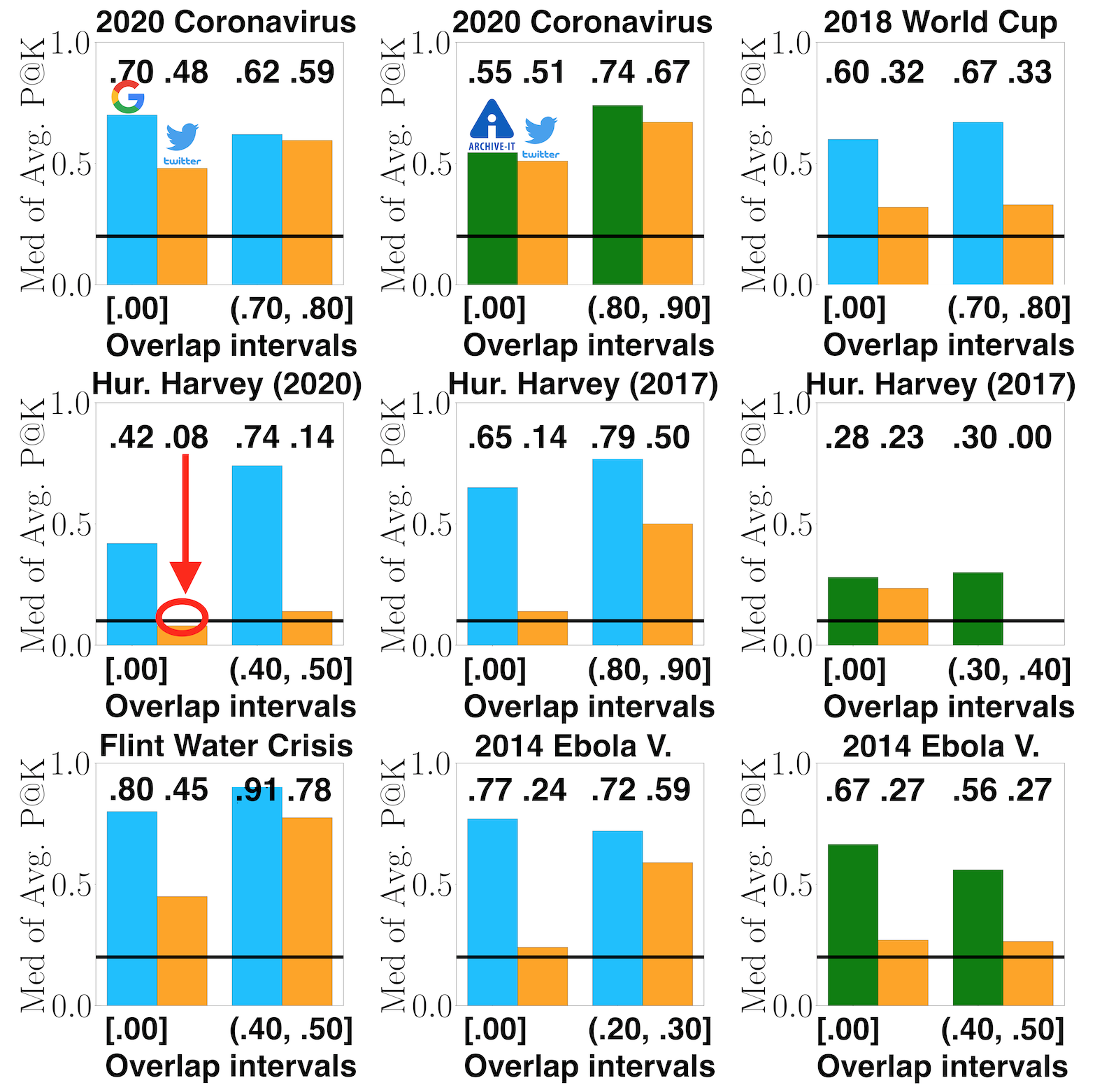} %
 
  \caption{Median of average P@K for minimum overlap ([0.00]) and maximum overlap (e.g., (.80, .90]) between top K QP seeds and reference Google/Expert seeds. The black line (0.20) marks the relevance threshold. In all cases except (red annotation) \textit{Hurricane Harvey} (collected 2020) the median of the average P@K of QP seeds for the 0 overlap (maximum novelty) interval was always above the relevance threshold.}
  \label{fig:nonNovel}
  
\end{figure}

\begin{table}[]
\centering
\caption{Combination of QPs that produced highest (\textcolor{blue}{QP Novel}) and lowest (\textcolor{red}{QP Non-Novel}) novelty. Non-novelty mostly favors seeds from broadly-reputable (e.g., $re_b$, $\overline{re_n}$) sources that are easy-to-find (e.g., $rt$, $\overline{sc}$) while novelty mostly favors seeds that are non-popular (e.g., non-popular author: $\overline{ap}$, non-popular posts: $\overline{rp}$, $\overline{sc}$), and hard-to-find (e.g., $\overline{rt}$)}
\setlength{\tabcolsep}{1.2pt}
\begin{tabular}{|c|c|c.c|c|}
\hline
\textbf{Dataset} & \multicolumn{2}{c.}{  \textbf{\textcolor{blue}{QP Novel}} } & \multicolumn{2}{c|}{  \textbf{\textcolor{red}{QP Non-Novel}} } \\ \cline{2-5}
 \textbf{Topic}  &    Google+Tw    &   Expert+Tw   &    Google+Tw         & Expert+Tw             \\ \Xhline{4\arrayrulewidth}
 Coronavirus &   $rl, \overline{ap}$    &   $rl, \overline{rp}$   & $ rl, re_b $    &   $ re_n $     \\ \hline
W. Cup ('18) &   $rl, \overline{sh}$    &         & $ ge_a, re_b, \overline{re_n} $ & \\ \hline
H. Harvey ('20) &   $rl, \overline{rt}$    &         & $ rt, re_b $  &      \\ \hline
H. Harvey ('17) &   $\overline{ap}, \overline{re_b}, \overline{re_n}$    &   $rl, \overline{rt}$   & $ rt, re_b $    &   $ \overline{sc} $    \\ \hline
Flint Water &   $rl, \overline{rp}, \overline{ap}$    &         & $ rt, \overline{re_b} $ & \\ \hline
Ebola V. ('14) &   $rl, \overline{rt}$    &   $rl, \overline{re_n}$   & $ rt, \overline{re_n} $   &   $ \overline{sc} $     \\ \hline
\end{tabular}
\label{tab:res013CombOvPrec}
\end{table}

\subsection{Results when Novelty \textcolor{red}{is not} prioritized}
\label{sec:resultsQualityNovel}
Consider the results (P@K/overlap) when novelty is not prioritized.
\subsubsection{\textbf{P@K of Twitter (and overlap with Google) QP seeds}}
Across all topics, for Google and Twitter seeds, the Minimum, Median, and Maximum (MMM) average overlap were \textcolor{gre}{\textbf{0.04}}, \textcolor{gre}{\textbf{0.32}}, and \textcolor{gre}{\textbf{1.0}}, respectively, when Quality Proxies were used to score seeds. Without the utilization of QP scores, the MMM average overlap were smaller, \textcolor{red}{\textbf{0.04}}, \textcolor{red}{\textbf{0.14}}, and \textcolor{red}{\textbf{0.27}}, respectively. These results (e.g., Fig. \ref{fig:qpImprovePrecOveralap}, right) suggest that the utilization of QP scores to rank and select seeds, helped surface seeds from a common set of domains between Twitter and Google unlike when QPs were not used. Additionally, they illustrate that different combinations of QP can result in high (e.g., $\boldsymbol{dp, ge_a, re_b}$ and $\boldsymbol{ge_d, re_b, \overline{re_n}}$) or low ($\boldsymbol{\overline{rl}, ge_d, \overline{re_n}}$) overlap/precision as expressed by Fig. \ref{fig:ovPrecScatt0}. In Fig. \ref{fig:ovPrecScatt0}, unsurprisingly, the QP combination $\boldsymbol{\overline{rl}, ge_d, \overline{re_n}}$ resulted in a low P@20 because the $relevance$ $rl$ QP was flipped, meaning relevance was penalized. Our results (Table \ref{tab:res013CombOvPrec}) also suggest that non-novelty mostly favors seeds from broadly-reputable (e.g., $re_b$, $\overline{re_n}$) sources that are easy-to-find (e.g., $rt$, $\overline{sc}$).

Across all topics, for Twitter seeds, with Google seeds as the reference, the MMM average Precision at K (P@K) were \textcolor{gre}{\textbf{0.0}}, \textcolor{gre}{\textbf{0.53}}, and \textcolor{gre}{\textbf{0.99}}, respectively, when QP scores were used. Without the utilization of QP scores, the MMM were smaller; \textcolor{red}{\textbf{0.06}}, \textcolor{red}{\textbf{0.45}}, and \textcolor{red}{\textbf{0.65}}, respectively. These results (e.g., Fig. \ref{fig:qpImprovePrecOveralap}, left) showed that the utilization of Quality Proxies to score, rank, and select seeds, improved the precision of seeds by 0.08 (0.53 vs 0.45).
\subsubsection{\textbf{P@K of Twitter (and overlap with Expert) QP seeds}}
\textcolor{white}{.}\\
Across all topics, for Expert and Twitter seeds, the MMM average overlap were \textcolor{gre}{\textbf{0.09}}, \textcolor{gre}{\textbf{0.67}}, and \textcolor{gre}{\textbf{1.0}}, respectively, when Quality Proxies were used to score and select seeds. Without the utilization of QPs, they were smaller, \textcolor{red}{\textbf{0.03}}, \textcolor{red}{\textbf{0.13}}, and \textcolor{red}{\textbf{0.19}}, respectively. Similar to the overlap between Google and Twitter seeds, these results suggest that the utilization of QP scores to rank and select seeds facilitated the selection of seeds from a common set of domains for Twitter and Expert seeds.

Across all topics, for Twitter seeds, with Expert seeds as the reference, the MMM average precision were \textcolor{gre}{\textbf{0.0}}, \textcolor{gre}{\textbf{0.72}}, and \textcolor{gre}{\textbf{0.95}}, respectively. Further investigation of the seeds that generated 0.0 precision showed that 5/10 were actually relevant based on human judgment. This means our relevance threshold of 0.20 was set too high, and thus resulted in the production of false positive labels. The MMM of the average precision of seeds not scored with QPs were smaller (\textcolor{red}{\textbf{0.06}}/\textcolor{red}{\textbf{0.55}}/\textcolor{red}{\textbf{0.71}}) by 0.17 (0.72 vs 0.55) suggesting again (as previously seen when Google was reference) that the utilization of QP scores improved the precision of seeds.
\subsection{Novelty \textcolor{blue}{is} prioritized: P@K of QP seeds}
\label{sec:resultsQualityNonNovel}
In Fig. \ref{fig:nonNovel}, the heights represent the median of the average P@K for different overlap intervals and horizontal lines mark the relevance threshold for each dataset topic. In all cases except \textit{Hurricane Harvey} (collected 2020), the median of the average P@K of Twitter QP seeds for the 0 overlap (maximum novelty) interval was always above the relevance threshold. This suggests that maximum novelty (0 overlap) did not adversely affect the P@K for Twitter QP seeds even though higher overlap resulted in a higher P@K. Our results (Table \ref{tab:res013CombOvPrec}) also suggest that novelty mostly favors seeds that are non-popular (e.g., non-popular author: $\overline{ap}$, non-popular posts: $\overline{rp}$, $\overline{sc}$), and hard-to-find (e.g., $\overline{rt}$).
\subsection{Correlation of Quality Proxies}
Our correlation analysis (Table \ref{tab:resCorrComp}) showed a strong ($>.50$) positive correlation between between popularity-based (e.g., $\{sh, lk, rp\}$) and reputation ($\{re_b, re_n\}$) QP metrics. All positive correlations were statistically significant (p <.05) unlike negative correlations. These results are not surprising. For example, a post with many likes ($lk$) is highly likely to be shared ($sh$) and/or replied to ($rp$). Similarly, many domains (e.g., \texttt{cdc.gov}) with high broad (topic) reputation ($re_b$) also have high narrow (story) reputation ($re_n$). 
\begin{table}[]
\centering
\caption{Pairs of QPs least and most correlated (Pearson's r) showing a strong positive correlation between popularity-based (e.g., $\{sh, lk, rp\}$) and reputation ($\{re_b, re_n\}$) QPs.
}
\setlength{\tabcolsep}{1.2pt}
\begin{tabular}{|c|c|c.c|c|}
\hline
\textbf{Rank} & \multicolumn{2}{c.}{  \textbf{Least Correlated} } & \multicolumn{2}{c|}{  \textbf{Most Correlated} } \\ \cline{2-5}
   &   \textbf{QPs}   &   \textbf{r} &    \textbf{QPs}         & \textbf{r}             \\ \Xhline{4\arrayrulewidth}
1. &   $sc, re_b$    &   -0.13 & $sh, lk$    & 0.94     \\ \hline
2. &   $dp, ge_a$    &   -0.05 & $re_b, re_n$& 0.82     \\ \hline
3. &   $ap, ge_d$    &   -0.04 & $rp, lk$    & 0.82     \\ \hline
4. &   $ge_a, rt$    &   -0.03 & $rp, sh$    & 0.82     \\ \hline
5. &   $ge_d, rt$    &   -0.03 & $dp, re_b$  & 0.66     \\ \hline
6. &   $ge_a, re_b$  &   -0.03 & $dp, re_n$  & 0.60     \\ \hline
7. &   $sc, re_n$    &   -0.02 & $ap, dp$    & 0.45     \\ \hline
\end{tabular}
\label{tab:resCorrComp}
\end{table}
\section{Future work and Conclusions}
The QP framework and metrics that instantiate the classes have some limitations for which a future work would address. First, the evaluation topics such as the \textit{2020 Coronavirus Pandemic} are well documented. We expect our framework to under-perform for esoteric or obscure stories due to sparse data. Second, the high correlation (Table \ref{tab:resCorrComp}) between QPs (e.g., popularity-based QPs $\{rp, sh, lk\}$) suggests popularity could be given more weight when combined with other QPs. Third, measuring relevance is limited by small text which could result in false negative errors.

The Web is one of the greatest outcomes of human endeavor, but it has some major
flaws, one of which is, the Web forgets, causing the disappearance of Web resources chronicling important stories and events. Web archive collections reduce this problem by preserving Web resources, and they begin with seed URLs hand-selected by experts or scraped from social media posts. While social media is a valuable source of seed URLs, the quality of social media content varies widely. In this paper, we presented the Quality Proxies framework (and instantiations) for assigning quality scores to seed URLs extracted from social media posts. A QP assigns a quality trait to a seed within a single dimension. Seeds can be assigned a quality score by selecting different combinations of Quality Proxies which map to different notions of quality across multiple dimensions such as $popularity$, $reputation$, \textit{geographical proximity}, etc. The QP framework is flexible (enables multiple definitions of quality), robust (operates with subsets), explainable, and extensible. Our results showed that Quality Proxies resulted in the selection of quality seeds with increased precision (by $\sim$0.13) when novelty is and is not prioritized. To encourage reproducibility we have provided our research data and code \cite{JCDL2021}.
\clearpage
\bibliographystyle{ACM-Reference-Format}
\balance
\bibliography{NwalaJCDL2021T} 
\end{document}